\documentclass[12pt]{article}

\usepackage{cite}
\usepackage{epsfig}
\usepackage{amsmath}
\usepackage{array}
\usepackage{pstricks}
\usepackage{bbold}

\def\mathswitch#1{\relax\ifmmode#1\else$#1$\fi}
\def\mathswitchr#1{\relax\ifmmode{\mathrm{#1}}\else$\mathrm{#1}$\fi}
\newcommand{\PW}{\mathswitchr W}
\newcommand{\PZ}{\mathswitchr Z}

\newcommand{\Pt}{\mathswitchr t}

\newcommand{\MW}{\mathswitch {M_\PW}}
\newcommand{\MZ}{\mathswitch {M_\PZ}}

\newcommand{\mt}{\mathswitch {m_\Pt}}

\newcommand{\anc}{\rule{0mm}{0mm}}

\newcommand{\mycaption}[1]{\caption{\sl #1}}

\begin{document}
\thispagestyle{empty}

\def\thefootnote{\fnsymbol{footnote}}

\begin{flushright}
\end{flushright}

\vspace{1cm}

\begin{center}

{\Large\sc {\bf Numerical evaluation of multi-loop integrals using subtraction
terms}}
\\[3.5em]
{\large\sc
A.~Freitas
}

\vspace*{1cm}

{\sl
Pittsburgh Particle-physics Astro-physics \& Cosmology Center
(PITT-PACC),\\ Department of Physics \& Astronomy, University of Pittsburgh,
Pittsburgh, PA 15260, USA
}

\end{center}

\vspace*{2.5cm}

\begin{abstract}
A formalism for the numerical integration of one- and two-loop integrals is
presented. It is based on subtraction terms which remove the soft, collinear and
some of the ultraviolet divergences from the integrand. The numerical integral
is performed in the Feynman parameter space, using a complex contour deformation
to ensure robust convergence even in the presence of physical
thresholds. The application of the proposed procedure is demonstrated with
several one- and two-loop examples. An implementation in the program {\sc
nicodemos} is publicly available, which currently incorporates only one-loop
functionality, but an extension to two-loop cases is planned for future
versions.
\end{abstract}

\setcounter{page}{0}
\setcounter{footnote}{0}

\newpage


\section{Introduction}

For many production processes at colliders and other particle physics
observables, the inclusion of higher-order radiative corrections is essential 
to match the experimental precision. Much progress towards efficient
calculational techniques has been made over the last decades by many authors.
However, the computation of loop diagrams with many different mass scales, many
external legs, or more than one loop remains a difficult task. In particular,
when going beyond the one-loop level, it is known that the loop integrals can in
general not be solved analytically. Nevertheless, by using specialized
semi-numerical techniques, several complete two-loop calculations have been
carried out, for example for electroweak precision observables \cite{ewprec,ew2}
and QCD corrections to the production  of gauge bosons \cite{lhcw}, top quarks
\cite{lhct} or Higgs bosons \cite{lhch} at the Large Hadron Collider
(LHC). However, the methods employed in these papers have been tailored to the
problem at hand and  cannot be applied easily to other situations.

Fully numerical techniques offer an alternative and potentially more flexible 
approach towards complex loop calculations. For a purely numerical method
to become viable, two main obstacles need to be overcome:
extracting the ultraviolet (UV) and infrared and
collinear (IR) singularities from the integral; and ensuring robust and efficient convergence of 
the numerical integration. Several powerful methods have been proposed:
\begin{itemize}
\item \emph{Sector decomposition} starts from the Feynman parametrization of a
loop diagram. It isolates the physical singularities through appropriate
mappings of the integration region \cite{sec}. There are additional integrable
singularities inside the integration region for a diagram with physical cuts.
While being formally integrable, such singularities cannot be handled by
standard numerical integration algorithms. To avoid the problem and improve numerical stability
one can deform the integration contours in the complex plane \cite{ns}.
\item The Feynman integrals can be transformed into \emph{Mellin-Barnes
representations}. An algorithm for the extraction of UV and IR poles has been
developed based the residue theorem \cite{mb,mb2}. Again, the presence of
physical cuts leads to bad convergence behavior of the numerical integrals,
which can be improved by variable mapping and contour deformations \cite{mbn}.
\item In a third approach the UV and IR singularities are removed from the
integrand by suitable {\emph subtraction terms} \cite{sub,sub2}. These subtraction
terms are designed such that they can be easily integrated analytically and
added back to the final result. As for the previous two methods, integrable 
singularities due to physical thresholds may be treated with the
help of contour deformations \cite{sub2}.
\end{itemize}
The method based on subtraction terms has so far been applied only to one-loop
calculations. This article presents a procedure for extending this idea to
two-loop integrals. For this purpose the IR subtraction terms are constructed in
analogy to Ref.~\cite{sub2,subd} and combined with the contour deformation prescription of
Ref.~\cite{ns}. In addition, a new treatment for the UV divergences is
introduced. As a first step, it is shown that this approach is already useful at
the one-loop level, leading to numerical integrals with good convergence
behavior. Secondly, it is demonstrated how it can be applied to two-loop
diagrams. In the present paper, only two-loop integrals with IR singularities in
one of the subloops are considered. The extension to overlapping IR
singularities in both subloops will be delegated to a forthcoming publication
\cite{subf}, but it is relatively straightforward since most of the required
subtraction terms can be constructed from the results of Ref.~\cite{ver2}.

A public computer code for the evaluation of arbitrary one-loop integrals is
provided, based on the method presented here. It is planned to incorporate
two-loop cases into future versions of this code.

In the next section, the method for the numerical evaluation of one-loop integrals,
using subtraction terms and contour deformation, is explained. It is
demonstrated how it can be cast into an algorithmic form and its application in
several examples is demonstrated. In section~\ref{sc:2} the extension to the
two-loop level is discussed and illustrated through a few examples. Finally, the
main results are summarized in
section~\ref{sc:sum}.


\section{One-loop integrals}
\label{sc:1}

A general one-loop integral, see Fig.~\ref{fig:oneloop}, may be written as
\begin{figure}[tb]
\centering
\pscircle[linewidth=1pt](0,-2){1}
\psline[linewidth=1pt](-2,-2)(-1,-2)\psdot(-1,-2)
\psline[linewidth=1pt](-1,-0.268)(-0.5,-1.134)\psdot(-0.5,-1.134)
\psline[linewidth=1pt](-1,-3.732)(-0.5,-2.866)\psdot(-0.5,-2.866)
\psline[linewidth=1pt](+1,-3.732)(+0.5,-2.866)\psdot(+0.5,-2.866)
\rput(-1.2,-3.9){$p_1$}\psline[linewidth=.5pt]{->}(-1.1,-3.682)(-0.9,-3.3356)
\rput(-2.8,-2){$p_2-p_1$}\psline[linewidth=.5pt]{->}(-2,-2.1)(-1.6,-2.1)
\rput(-1.0,-0.1){$p_3-p_2$}\psline[linewidth=.5pt]{->}(-1.1,-0.318)(-0.9,-0.6464)
\psdot[dotsize=1pt](0,-0.6)
\psdot[dotsize=1pt](0.7,-0.7876)
\psdot[dotsize=1pt](1.2124,-1.3)
\psdot[dotsize=1pt](1.4,-2)
\psdot[dotsize=1pt](1.2124,-2.7)
\rput(+1.4,-3.9){$p_n-p_{n-1}$}\psline[linewidth=.5pt]{->}(+1.1,-3.682)(+0.9,-3.3356)
\rput[t](0,-3.1){$m_0$}
\rput[r](-0.9,-2.6){$m_1$}
\rput[r](-0.9,-1.4){$m_2$}
\rput(0,-2.6){$k$}\psline[linewidth=.5pt]{->}(-0.2,-2.85)(0.2,-2.85)
\anc\\[3.7cm]
\mycaption{Topology of a general one-loop integral.
\label{fig:oneloop}}
\end{figure}
\begin{align}
&I^{(1)} = \int \widetilde{dk} \; \frac{N(k)}{D^{(1)}(k)}, \qquad
\widetilde{dk} = e^{\gamma_{\rm E}(4-d)/2}\frac{d^dk}{i\pi^{d/2}},
\\
&D^{(1)}(k) = [k^2-m_0^2][(k-p_1)^2-m_1^2]
\cdots [(k-p_n)^2-m_n^2], \label{eq:i1}
\end{align}
where $N(k)$ is a polynomial in the loop momentum $k$. This integral may have
soft, collinear and ultraviolet divergences. The first two are canceled in the
integrand by using suitable subtraction terms, similar to Ref.~\cite{sub2} (see
also Ref.~\cite{subd}).
However, no UV subtraction terms are used for the one case here; instead the UV
poles are evaluated explicitly after introducing Feynman parameters.


\subsection{Infrared subtraction terms}
\label{sc:irsub}

The $i$th propagator of integral \eqref{eq:i1} produces a soft singularity if
$m_i=0$, $(p_{i+1}-p_i)^2=m_{i+1}^2$, $(p_{i-1}-p_i)^2=m_{i-1}^2$, and $N(k) \neq 0$ for
$k\to p_i$. The soft subtraction term is given by \cite{sub2,subd}
\begin{align}
G^{(1)}_{\rm soft} &= \frac{1}{[(k-p_{i-1})^2-m_{i-1}^2][(k-p_i)^2]
 [(k-p_{i+1})^2-m_{i+1}^2]} \, N(k{=}p_i)\hspace{-.7em} 
 \prod_{\substack{j\neq \\  i-1,\,i,\,i+1}}\hspace{-.7em}
 \frac{1}{(p_i-p_j)^2-m_j^2} \nonumber \\
&= \frac{1}{D^{(1)}(k)} \, N(k{=}p_i)\hspace{-.7em} 
 \prod_{\substack{j\neq \\  i-1,\,i,\,i+1}}\hspace{-.7em}  
 \frac{(k-p_j)^2-m_j^2}{(p_i-p_j)^2-m_j^2},
 \label{eq:is}
\end{align}
where, for later convenience, the last line has been written in terms of the common denominator
of eq.~\eqref{eq:i1}. The integrated soft subtraction term reads
\begin{align}
&m_{i-1}=m_{i+1}=0: \quad
\int \widetilde{dk} \; G^{(1)}_{\rm soft} = 
  F_{i,\rm rem}\,\frac{1}{s}\biggl [ \frac{1}{\varepsilon^2} - \frac{1}{\varepsilon}\log(-s)
  + \frac{\log^2(-s)}{2} - \frac{\pi^2}{12} \biggr ], \\[1ex]
&m_{i-1}>0,\,m_{i+1}=0: 
\int \widetilde{dk} \; G^{(1)}_{\rm soft} = \begin{aligned}[t] 
  &F_{i,\rm rem}\,\frac{1}{s-m_{i-1}^2}\biggl [ \frac{1}{2\varepsilon^2} -
   \frac{1}{\varepsilon}\log\Bigl (\frac{m_{i-1}^2-s}{m_{i-1}}\Bigr ) 
   +\frac{\pi^2}{24} \\ &+ \frac{\log^2(m_{i-1}^2-s)}{2} - \log^2(m_{i-1}) 
   -\text{Li}_2\Bigl (\frac{-s}{m_{i-1}^2-s}\Bigr ) \biggr ],
 \end{aligned} \\
&m_{i-1},\,m_{i+1}>0: \nonumber \\
&\quad 
\begin{aligned}[c]
 \int\widetilde{dk}\;&G^{(1)}_{\rm soft} = F_{i,\rm rem}\,
  \frac{-x}{m_{i-1}m_{i+1}(1-x^2)} \\
 &\times \biggl [ \log(x)\, \Bigl (
  -\frac{1}{\varepsilon} - \frac{\log(x)}{2} + 2 \log(1-x^2) +
  \log(m_{i-1}m_{i+1})\Bigr ) \\
 &\;\;\;\;-\frac{\pi^2}{6} + \text{Li}_2(x^2) +
 \frac{1}{2}\log\Bigl (\frac{m_{i-1}}{m_{i+1}}\Bigr ) + 
 \text{Li}_2\Bigl (1+x\frac{m_{i-1}}{m_{i+1}}\Bigr ) + 
 \text{Li}_2\Bigl (1+x\frac{m_{i+1}}{m_{i-1}}\Bigr )
 \biggr ],
\end{aligned} \\[.5ex]
&\qquad x = -\frac{1-\xi}{1+\xi}+i\epsilon, \qquad
\xi = \sqrt{1-\frac{4m_{i-1}m_{i+1}}{s-(m_{i-1}-m_{i+1})^2}},
\end{align}
where $\varepsilon=(4-d)/2$, $s=(p_{i+1}-p_{i-1})^2+i\epsilon$, and
\begin{equation}
F_{i,\rm rem} = N(k{=}p_i)\hspace{-.7em} 
 \prod_{\substack{j\neq \\  i-1,\,i,\,i+1}}\hspace{-.7em}
 \frac{1}{(p_i-p_j)^2-m_j^2}.
\end{equation}
A collinear singularity singularity is encountered if $m_i=m_{i-1}=0$,
$(p_i-p_{i-1})^2=0$, and $N(k) \neq 0$ for $k\to p_i$. The simplest subtraction
term for this singularity is given by
\begin{align}
G^{(1)}_{\rm coll} &= \frac{1}{(k-p_{i-1})^2(k-p_i)^2} \, 
N(k{=}p_i)\hspace{-.5em} 
 \prod_{j\neq i-1,\,i} \frac{1}{(p_i-p_j)^2-m_j^2} \nonumber \\
&= \frac{1}{D^{(1)}(k)} \, N(k{=}p_i)\hspace{-.5em} 
 \prod_{j\neq i-1,\,i} \frac{(k-p_j)^2-m_j^2}{(p_i-p_j)^2-m_j^2}.
\end{align}
In dimensional regularization, the integrated collinear subtraction term is
simply zero,
\begin{equation}
\int \widetilde{dk} \; G^{(1)}_{\rm coll} = 0.
\end{equation}
In other words, by subtracting $G^{(1)}_{\rm coll}$ from the integrand the
collinear singularity is transformed into a UV singularity, which will be
handled as explained below.
Note that this collinear subtraction term only works when applied to
physical amplitude, $i.\,e.$ to a gauge-invariant set of diagrams
\cite{sub2}.


\subsection{Variable mapping and contour deformation}
\label{sc:def}

After subtracting the IR singularities as described above, one
arrives at a loop integral that only contains UV divergences,
\begin{equation}
I^{(1)}_{\rm reg} \equiv I^{(1)} - \sum \int \widetilde{dk}\, G^{(1)}_{\rm soft} 
 - \sum \int \widetilde{dk}\, G^{(1)}_{\rm coll},
\end{equation}
where the sums are over all soft and collinear singularities in $I^{(1)}$.
It has a structure similar to eq.~\eqref{eq:i1}:
\begin{equation}
I^{(1)}_{\rm reg} = \int \widetilde{dk} \; 
 \frac{N_{\rm reg}(k)}{D^{(1)}(k)}.
\end{equation}
By introducing Feynman parameters and shifting the loop momentum, 
this expression can be cast into the form
\begin{equation}
I^{(1)}_{\rm reg} = \int dx_1\dots dx_n \; \delta(1-{\textstyle\sum} x_i) 
 \int \widetilde{dk} \;
 \frac{\tilde{N}(k)}{[k^2-A]^n}, \label{eq:if}
\end{equation}
where $\tilde{N}(k)$ is a polynomial in $k$ and in the Feynman parameters
$x_1, ..., x_n$, while $A$ is a polynomial in $x_1, ..., x_n$. 
It is convenient to map the Feynman parameters onto a hypercube:
\begin{align}
&\begin{aligned}
x_1 &= 1-y_1,\\
x_2 &= y_1(1-y_2),\\
&\;\;\vdots \\
x_{n-1} &= y_1\cdots y_{n-2}(1-y_{n-1}), \\
x_n &= y_1\cdots y_{n-2}y_{n-1},
\end{aligned}
&&
\begin{aligned}
\int &dx_1\dots dx_n \; \delta(1-{\textstyle\sum} x_i) \\
&= \int_0^1 dy_1\dots dy_n \; y_1^{n-2}y_2^{n-3}\cdots y_{n-2}.
\end{aligned}
\end{align}
The $k$-dependence in the numerator of \eqref{eq:if} is eliminated by using the
tensor reduction formula
\begin{equation}
\begin{aligned}
\int\widetilde{dk} \; &\frac{k^{\mu_1}k^{\mu_2}\cdots k^{\mu_r}}{[k^2-A]^n}
\\
 &= \frac{1}{r!!\,d(d+2)\cdots(d+r-2)}\sum_{\rm permut.} (g^{\mu_1\mu_2} \cdots g^{\mu_{r-1}\mu_r})
  \int\widetilde{dk} \; \frac{k^r}{[k^2-A]^n},
 \label{eq:ten1}
\end{aligned}
\end{equation}
where the sum is over all permutations of $\mu_1, ...,\mu_r$. Thus one arrives at
\begin{equation}
I^{(1)}_{\rm reg} = \int_0^1 dy_1\dots dy_{n-1} 
 \int \widetilde{dk} \;
 \biggl [ \frac{C_1}{[k^2-A]^n} + \frac{C_2}{[k^2-A]^{n-1}} + \dots \biggr ],
 \label{eq:it}
\end{equation}
where the $C_i$ are polynomials in $y_1, ..., y_n$ but independent of $k$.
Integrating over $k$, one obtains
\begin{equation}
\begin{aligned}
I^{(1)}_{\rm reg} = \int_0^1 dy_1\dots dy_{n-1}
 \biggl [ &C_1(-1)^n\frac{\Gamma(n+\varepsilon-2)}{\Gamma(n)} A^{-n-\varepsilon+2}
 \\
  + &C_2(-1)^{n-1}\frac{\Gamma(n+\varepsilon-3)}{\Gamma(n-1)} A^{-n-\varepsilon+3}
  + \dots \biggr ],
 \label{eq:int1}
\end{aligned}
\end{equation}
which can be expanded in $\varepsilon$, so that the UV singularities appear as
$1/\varepsilon$ poles. Thus the integral takes the form
\begin{equation}
I^{(1)}_{\rm reg} = \int_0^1 dy_1\dots dy_{n-1}
 \biggl [ D_0\Bigl (\frac{1}{\varepsilon} + \log(A-i\epsilon)\Bigr )
  + D_1 (A-i\epsilon)^{-1} + D_2 (A-i\epsilon)^{-2}
  + \dots \biggr ], \label{eq:ie}
\end{equation}
where the $i\epsilon$ from the propagators have been made explicit again. $A$
and the $D_i$ are polynomials in the variables $y_1, ...,y_{n-1}$.

The coefficients in the $\varepsilon$-expansion in eq.~\eqref{eq:ie} are finite,
so that the integration over the parameters $y_1, ...,y_{n-1}$ can be performed
numerically. However, the polynomial $A(\vec{y})$ can have zeros inside
the integration region, which happens 
if the loop integral has physical thresholds. While these singularities are formally
integrable, they lead to problems for the numerical integration. A solution is
the deformation of the integration contours into the complex plane, by using the
variable transformation \cite{ns}
\begin{equation}
y_i = z_i - i\lambda z_i(1-z_i)\frac{\partial A}{\partial z_i}, \qquad
0 \leq z_i \leq 1. \label{eq:cd}
\end{equation}
To leading order in $\lambda$ this produces a negative imaginary part in $A$:
\begin{equation}
A(\vec{y}) = A(\vec{z}) - i\lambda \sum_i z_i(1-z_i)
 \Bigl (\frac{\partial A}{\partial z_i}\Bigr )^2 + {\cal O}(\lambda^2).
\end{equation}
So if $\lambda$ is chosen small enough, the integral
\begin{equation}
I^{(1)}_{\rm reg} = \int_0^1 dz_1\dots dz_{n-1} \, \left|
 \frac{\partial(y_1, \dots, y_{n-1})}{\partial(z_1, \dots, z_{n-1})}\right| \,
 \biggl [ D_0\Bigl (\frac{1}{\varepsilon} + \log(A-i\epsilon)\Bigr )
  + D_1 A^{-1} + D_2 A^{-2}
  + \dots \biggr ]
\end{equation}
is well-behaved for numerical integration. In most cases, a good choice for
$\lambda$ is roughly 0.5.


\subsection{Numerical examples}

To demonstrate the flexibility and efficiency of the method described in the
previous section, several example calculations have been carried out and, where
applicable, compared to existing analytical results.

The numerical integration has been performed with the {\sc Vegas} and {\sc
Cuhre} algorithms of the {\sc Cuba 1.4} library \cite{cuba}. Timing information
is given for running on a single core of a {\sc
Intel\textsuperscript{\textregistered} Xeon\textsuperscript{\textregistered}
X5570} processor with 2.93~GHz.

\begin{figure}[t]
\centering
\psfig{figure=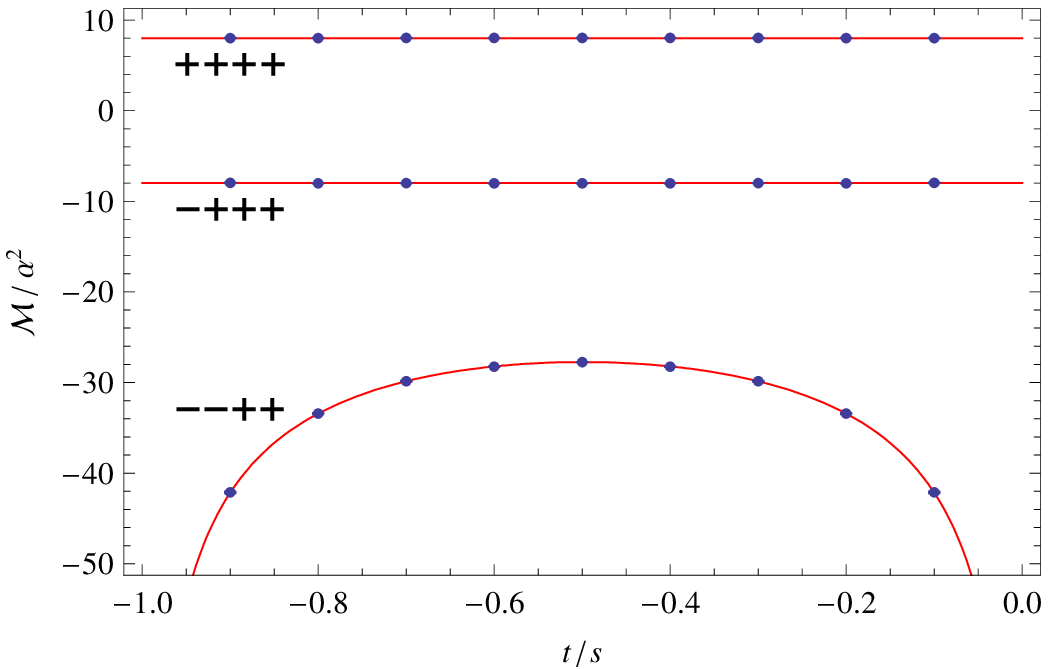, width=5in}
\vspace{-2ex}
\mycaption{Results for the four-photon amplitude ${\cal M}(\gamma\gamma \to
\gamma\gamma)$ (normalized relative to $\alpha^2$) for different photon helicity
combinations. The dots correspond to the results obtained with the numerical
method presented here, while the lines depict the analytical result from Ref.~\cite{Binoth:2002xg}.
The contour deformation parameter has been chosen $\lambda=1$.
\label{fig:gggg}}
\end{figure}

\paragraph{1.}
Let us start with the amplitude for two-photon scattering $\gamma\gamma \to
\gamma\gamma$ at the one-loop level, with electrons running in the loop. This
process does not involve any physical singularities and thus is a simple test of
the contour deformation and numerical integration. The six contributing diagrams
are grouped into three groups, according to the structure of the loop
denominators. Each group is evaluated separately, using the {\sc Vegas} algorithm with 
$10^6$ integration points each, which in total takes 4.8~s evaluation time.
Results for different
helicity combinations of the photons are depicted in Fig.~\ref{fig:gggg}, and
compared to analytical results from Ref.~\cite{Binoth:2002xg}.
The numerical integration error ranges between 0.2\% and 0.5\%;
error bars are included but not visible in the plot.

\begin{figure}[tb]
\vspace{.5ex}
\psfig{figure=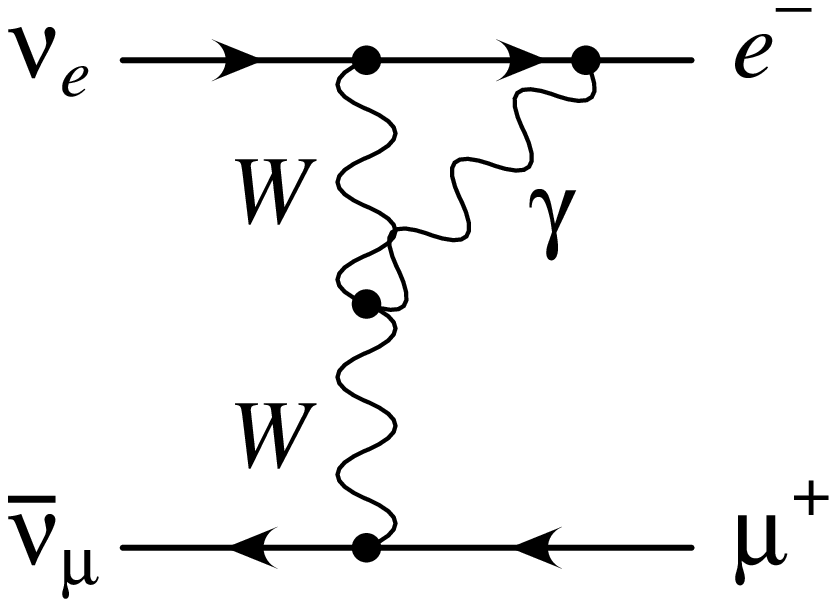, height=1in}
\hfill
\psfig{figure=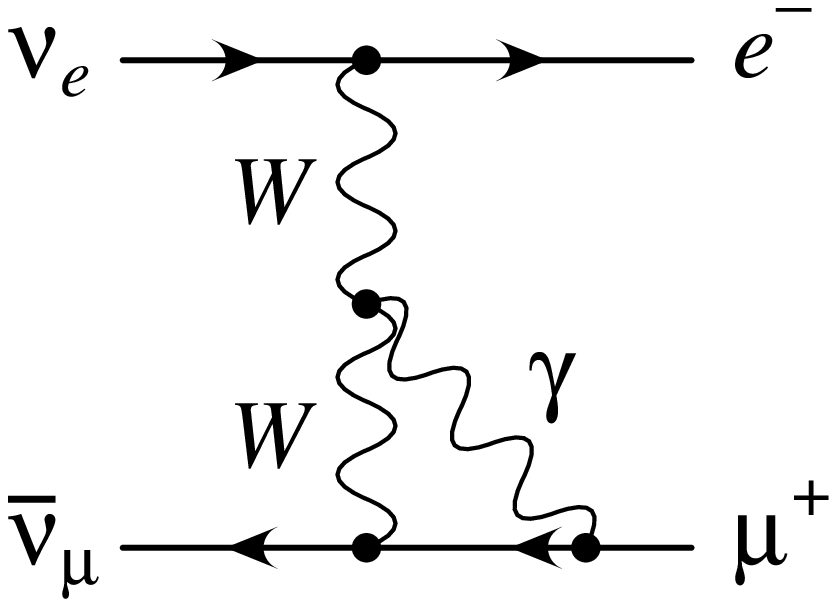, height=1in, bb=230 307 410 490, clip=true}
\hfill
\psfig{figure=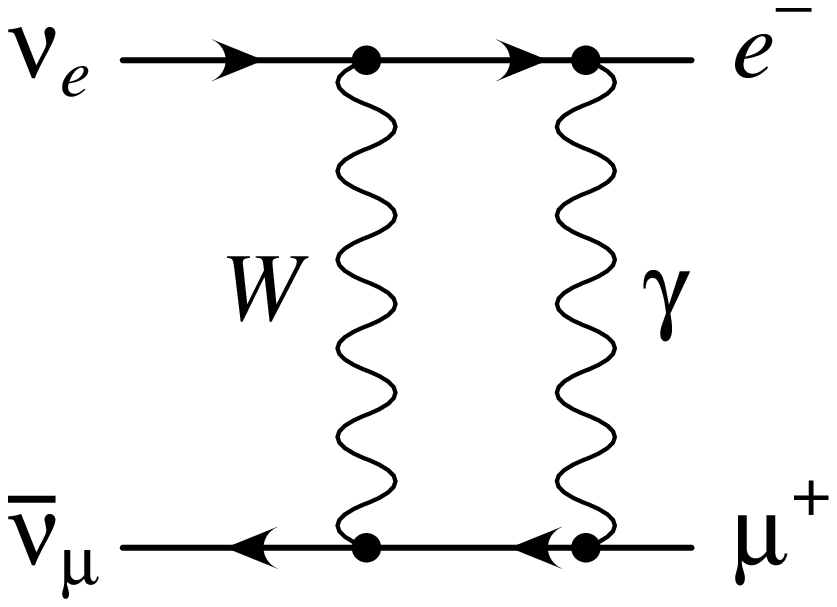, height=1in, bb=230 307 410 490, clip=true}
\hfill
\psfig{figure=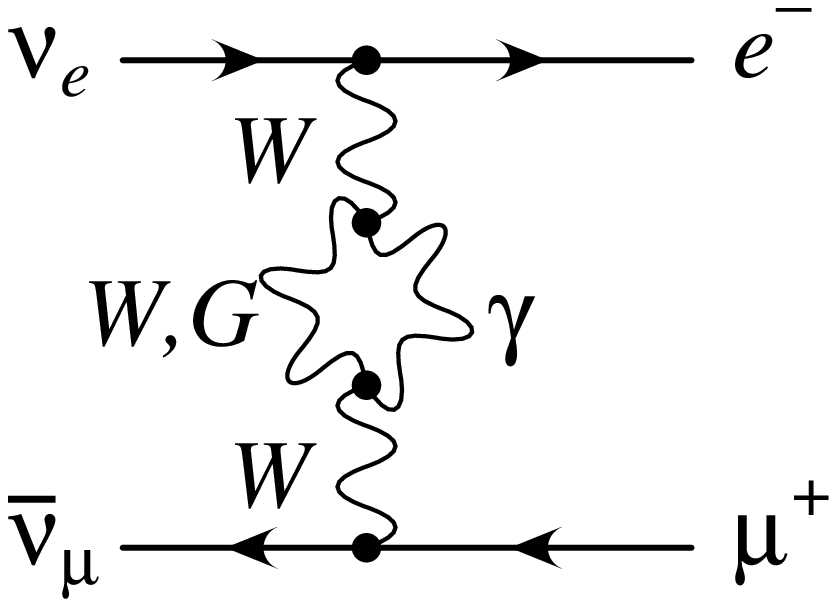, height=1in, bb=229 307 410 490, clip=true}
\hfill
\psfig{figure=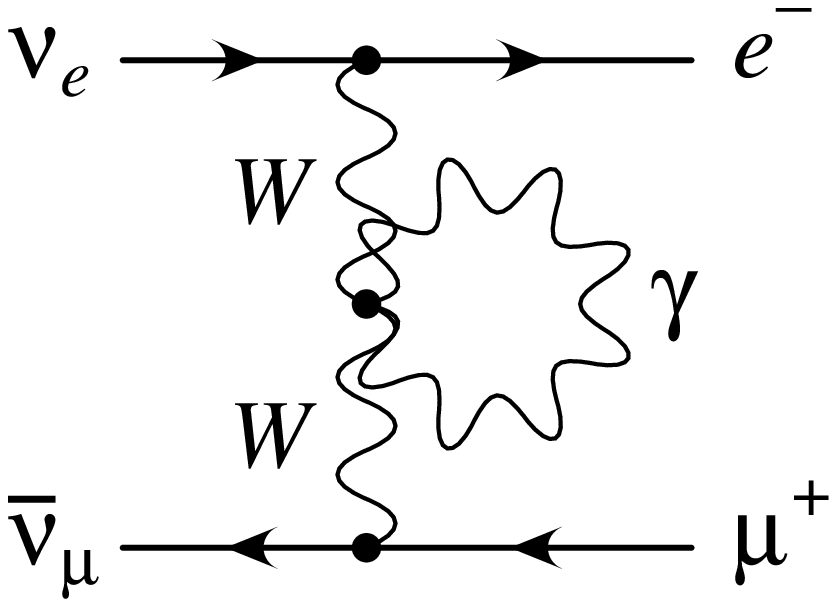, height=1in, bb=230 307 410 490, clip=true}
\vspace{-2ex}
\mycaption{Feynman diagrams for the one-loop QED corrections to
$\nu_e\bar{\nu}_\mu \to e^-\mu^+$.
\label{fig:nnll}}
\end{figure}

\paragraph{2.}
The one-loop QED corrections to $\nu_e\bar{\nu}_\mu \to e^-\mu^+$ is an example
with UV, soft and collinear singularities. The six diagrams in
Fig.~\ref{fig:nnll} are combined into a single expression with one common
denominator, from which the soft and collinear divergences are subtracted as
explained in section~\ref{sc:irsub}.

The results presented in Tab~\ref{tab:nnll} correspond to the next-to-leading
order correction to the squared matrix element, given by $2\, \text{Re}\{{\cal
M}^{(0)}{\cal M}^{(1)}\}$, where ${\cal M}^{(0)}$ and ${\cal M}^{(1)}$ refer to
the tree-level and one-loop QED corrected matrix elements, respectively.
The numerical integration has been performed with the {\sc Cuhre} algorithm,
using $5\times 10^5$ integration points (resulting in 2.4~s running time). 
Note that the $\varepsilon^{-2}$ pole stems
solely from the soft subtraction term and thus is known analytically.
\begin{table}
\centering
\renewcommand{\arraystretch}{1.2}
\begin{tabular}{lll}
\hline
 & \multicolumn{1}{c}{this work} & \multicolumn{1}{c}{analytical} \\
\hline
${\cal O}(\varepsilon^{-2})$ & $-$0.625 & $-$0.625 \\
${\cal O}(\varepsilon^{-1})$ & $\phantom{-}1.13113365-1.96349541i$ & 
  $\phantom{-}1.13113365 - 1.96349541i$ \\
${\cal O}(\varepsilon^{0})$ & $\phantom{-}3.27791(4)+1.83014(4)i$ & 
  $\phantom{-}3.27792343 + 1.83014477i$ \\
\hline
\end{tabular}
\mycaption{Numerical results for the next-to-leading order QED corrections to
the squared matrix element of $\nu_e\bar{\nu}_\mu \to e^-\mu^+$. Specifically,
the table shows numbers for $2\, \text{Re}\{{\cal M}^{(0)}{\cal M}^{(1)}\}$,
for the input parameters $\MW = 1$, $s=2$, $t=-1$, and $\lambda=0.5$. The results from the
numerical method described in this paper are given in the second column,
together with the numerical error from $5\times 10^5$ integration points of the {\sc
Cuhre} integration routine. For
comparison the third column lists the results obtained with the traditional
analytical approach using Passarino-Veltman reduction.
\label{tab:nnll}}
\end{table}
For comparison, the table also shows the result from a traditional analytical
calculation using Passarino-Veltman reduction \cite{pv} and explicit expression
for the well-known basic integrals, see $e.\,g.$ Ref.~\cite{oneloop}.

\paragraph{3.}
A third example, with a more difficult denominator structure, is the scalar
hexagon integral depicted in Fig.~\ref{fig:hex}. It contains several mass scales, has
three soft singularities, and requires contour deformation. Numerical results
are shown for the input values
\begin{equation}
\begin{aligned}
&m_{\rm a}^2 = 1.0, &\qquad & \vec{p}_1 = (0,0,3), && \vec{p}_2 = (0,0,-3), \\
&m_{\rm b}^2 = 0.25, && \vec{p}_3 = (0.5,0,0.5), && \vec{p}_4 = (-0.2,0,0.1), \\
&m_{\rm c}^2 = 4.0, && \vec{p}_5 = (0,1.37626,-0.3), && \vec{p}_6 =
 (-0.3,-1.37626,-0.3),\\
\end{aligned}
\end{equation}
and $\lambda=0.5$. Using $10^7$ integration points with the {\sc Cuhre}
algorithm one obtains
\begin{equation}
\begin{aligned}
&{\cal O}(\varepsilon^{-1}): && -0.0044718804 + 0.0120697975 i, \\
&{\cal O}(\varepsilon^{0}): && -0.04383(14) + 0.00790(14) i,
\end{aligned}
\end{equation}
which takes 45~min to evalute on the test computer used here.
\begin{figure}[tb]
\centering
\psline[linewidth=1pt](0,-0.5)(1,-1)(1,-2)(0,-2.5)(-1,-2)(-1,-1)(0,-0.5)
\psline[linewidth=1pt](-2,-1)(-1,-1)
\psline[linewidth=1pt](-2,-2)(-1,-2)
\psline[linewidth=1pt](2,-1)(1,-1)
\psline[linewidth=1pt](2,-2)(1,-2)
\psline[linewidth=1pt](1.5,-0.5)(0,-0.5)
\psline[linewidth=1pt](1.5,-2.5)(0,-2.5)
\psdot(-1,-1)
\psdot(-1,-2)
\psdot(0,-0.5)
\psdot(0,-2.5)
\psdot(1,-1)
\psdot(1,-2)
\rput(-2,-0.85){$p_1 \to \phantom{|}$}
\rput(-2,-2.15){$p_2 \to \phantom{|}$}
\rput(1.5,-0.35){$\phantom{|} \to p_3$}
\rput(2,-0.85){$\phantom{|} \to p_4$}
\rput(2,-2.15){$\phantom{|} \to p_5$}
\rput(1.5,-2.65){$\phantom{|} \to p_6$}
\rput[t](-1.5,-1.1){$m_{\rm a}$}
\rput[b](-1.5,-1.9){$m_{\rm a}$}
\rput[l](-0.9,-1.5){$m_{\rm a}$}
\rput[rb](-0.6,-0.7){0}
\rput[rt](-0.6,-2.3){0}
\rput[rt](0.5,-0.8){$m_{\rm b}$}
\rput[rb](0.5,-2.2){$m_{\rm c}$}
\rput[t](1.5,-1.1){$m_{\rm b}$}
\rput[b](1.5,-1.9){$m_{\rm c}$}
\rput[b](0.7,-0.4){$m_{\rm b}$}
\rput[t](0.7,-2.6){$m_{\rm c}$}
\rput[r](0.9,-1.5){0}
\anc\\[2.7cm]
\mycaption{Scalar hexagon diagram with soft singularities, labeling the masses
of all internal and external lines and the external momenta.
\label{fig:hex}}
\end{figure}


\subsection{The computer package {\sc nicodemos}}
\label{sc:nico}

The procedure outlined in sections~\ref{sc:irsub} and \ref{sc:def} has been
implemented into the public computer code {\sc nicodemos} (Numerical Integration
with COntour DEformation and MOdular Subtractions). The package contains a {\sc
Mathematica} module, which performs the application of subtraction terms and
Feynman parametrization (with or without deformation). The user needs to supply 
the input expression and the location of IR singularities. The {\sc
Mathematica} code then produces a {\sc Fortran} executable, which performs the
numerical integration for a given set of numerical input parameters.

{\sc nicodemos} is available for download at {\tt http://www.pitt.edu/\~{}afreitas/} and
can be used freely, provided the source is acknowledged and properly cited.
Version 1.0 can only handle one-loop integrals, but it is planned to expand the
functionality to include two-loop integrals in future versions.


\section{Two-loop integrals}
\label{sc:2}

In this section the extension to two-loop diagrams is discussed. The two-loop
integrals may contain UV singularities as well as IR singularities in one
subloop, while the case with IR singularities in both subloops will be
addressed in a future publication \cite{subf}.

Of particular interest are relatively complex two-loop diagrams, $i.\,e.$
diagrams with a relatively large number of external legs. Therefore the
following discussion will not consider
some special cases that occur only for two-loop tadpole and selfenergy diagrams,
since these can be evaluated with existing methods, see $e.\,g.$ Ref.~\cite{2self}.

Following the notation of the previous section, a two-loop integral is given by
\begin{align}
&I^{(2)} = \int \widetilde{dk}_1 \widetilde{dk}_2 \; 
\frac{N(k_1,k_2)}{D^{(2)}(k_1,k_2)},
\\
&D^{(2)}(k_1,k_2) = \begin{aligned}[t]
&[k_1^2-m_0^2][(k_1-p_1)^2-m_1^2] \cdots [(k_1-p_r)^2-m_r^2] \\
&\times [(k_2-p_{r+1})^2-m_{r+1}^2] \cdots [(k_2-p_s)^2-m_s^2]\\ 
&\times [(k_1-k_2-p_{s+1})^2-m_{s+1}^2] \cdots 
 [(k_1-k_2-p_n)^2-m_n^2],
\end{aligned} 
\label{eq:i2}
\end{align}
where $N(k_1,k_2)$ is polynomial in $k_1$ and $k_2$ and in the external momenta.

If one of the two subloops has an IR singularity it can be subtracted analogously
to the one-loop case. For example, the scalar diagram in Fig.~\ref{fig:ver2a}
\begin{equation}
I_{\rm fig \ref{fig:ver2a}} \begin{aligned}[t] = 
 &\int \widetilde{dk}_1 \widetilde{dk}_2 \\[-1ex]
 \times\, &\frac{1}{[k_1-m^2][(k_1-p_1)^2][(k_1-p_1-p_2)^2-m^2][(k_2-p_1)^2-m^2]
  [(k_1-k_2)^2-m^2]}
\end{aligned}
\end{equation}
has a soft singularity for $k_1\to p_1$.
\begin{figure}[tb]
\centering
\psline[linewidth=1pt](-2,-2)(-1,-2)(1,-0.8)(1,-2)
\psline[linewidth=1pt](-1,-2)(1.75,-3.65)
\psline[linewidth=1pt](1,-0.8)(1.75,-0.35)
\pscircle[linewidth=1pt](1,-2.56){0.56}
\psdot(-1,-2)
\psdot(1,-2)
\psdot(1,-0.8)
\psdot(0.75,-3.05)
\rput[r](-1.8,-2.2){$p_1+p_2 \to$}
\rput[r](-1.8,-2.8){$[(p_1+p_2)^2=s]$}
\rput{34}(1.8,-0.55){$\to$}
\rput[l](2.1,-0.2){$p_1 \quad [p_1^2=m^2]$}
\rput{-34}(1.8,-3.45){$\to$}
\rput[l](2.1,-3.6){$p_2 \quad [p_2^2=m^2]$}
\rput[r](0.5,-2.1){$m$}
\rput[l](1.65,-2.6){$m$}
\rput[rb](0,-1.3){$m$}
\rput[tr](0,-2.7){$m$}
\rput[l](1.1,-1.4){0}
\anc\\[3.7cm]
\mycaption{Scalar two-loop diagram with soft singularity. The figures also
labels the masses of all internal and external lines and the external momenta.
\label{fig:ver2a}}
\end{figure}
The subtraction term is constructed as in eq.~\eqref{eq:is}, \emph{viz.}
\begin{equation}
G^{\rm fig \ref{fig:ver2a}}_{\rm soft} = 
 \frac{1}{[k_1-m^2][(k_1-p_1)^2][(k_1-p_1-p_2)^2-m^2][(k_2-p_1)^2-m^2]
  [(p_1-k_2)^2-m^2]}.
\label{eq:ss2}
\end{equation}
This expression factorizes into two one-loop integrals, which can be integrated
analytically.

In the same way, one can subtract collinear divergences in one subloop with the
subtraction terms introduced in section~\ref{sc:irsub}.


\subsection{Ultraviolet divergences}

Since two-loop integrals may have overlapping UV singularities from both
subloops, the corresponding $1/\varepsilon$ poles cannot be computed directly as
in eq.~\eqref{eq:ie}. Instead one has to introduce subtraction terms also for the
UV divergences. The UV subtraction is performed in two steps:

\paragraph{1.}
The global UV singularities of both subloops can be obtained by performing a
Taylor expansion of the two-loop amplitude in terms of the external momenta. For
all physical amplitudes with three or more external legs only the leading term
in this expansion contributes to the global UV divergence. Two-loop tadpoles and
selfenergies are not considered here, as mentioned above. Thus the global UV
subtraction term is defined as
\begin{equation}
G^{(2)}_{\rm glob} = \frac{N(k_1,k_2)}{D^{(2)}(k_1,k_2)}\bigg|_{p_i=0}
\end{equation}
The integrated subtraction term $\int \widetilde{dk}_1 \widetilde{dk}_2 \;
G^{(2)}_{\rm glob}$ consists of two-loop vacuum integrals, which can be
evaluated analytically with the methods of Ref.~\cite{Davydychev:1992mt}.

\paragraph{2.}
The remainder $I^{(2)}_{\rm gs} \equiv I^{(2)}_{\rm reg} - \int \widetilde{dk}_1
\widetilde{dk}_2 \; G^{(2)}_{\rm glob}$ can still contain a UV singularity in
one of the subloops, or in both. The latter case, however, only occurs for
tadpole and selfenergy diagrams and thus will not be considered here, since it can
be handled more efficiently with other methods \cite{2self}.

Let us then assume that only the subloop with loop momentum $k_1$ has a 
UV divergence. The $k_1$ integral is now turned into a Feynman-parameter
integral by following the steps in eqs.~\eqref{eq:if}--\eqref{eq:it}, leading to
\begin{equation}
I^{(2)}_{\rm gs} = \int_0^1 dy_1\dots dy_{m-1} 
 \int \widetilde{dk_1}\widetilde{dk_2} \;
 \biggl [ \frac{C_1}{[k_1^2-A]^m} + \frac{C_2}{[k_1^2-A]^{m-1}} + \dots \biggr ],
\label{eq:it2}
\end{equation}
where the $C_i$ and $A$ depend on $k_2$ and the Feynman parameters $y_1, ...,
y_{m-1}$. Here $m=n+r-s$ is the number of propagators with $k_1$. 
Note that the $C_i$ are rational functions containing the propagators that
depend on $k_2$ only. At this point, a
subloop UV singularity is indicated by a term of the form $C_j/[k_1^2-A]^2$
in eq.~\eqref{eq:it2}. All higher powers of $[k_1^2-A]$ in the denominator are
UV-finite. The subloop UV divergence can be subtracted by
\begin{equation}
G^{(2)}_{\rm sub} = \int_0^1 dy_1\dots dy_{m-1}\, \frac{C_j}{[k_1^2-\mu^2]^2},
\end{equation}
where $\mu$ is a suitably chosen mass parameter. The integrated subtraction term
factorizes into two one-loop integrals:
\begin{equation}
\int \widetilde{dk}_1\widetilde{dk}_2 \, G^{(2)}_{\rm sub} = 
 -\Gamma(\varepsilon-2)\mu^{2-\varepsilon} \int_0^1 dy_1\dots dy_{m-1} 
 \int\widetilde{dk}_2 \, C_j.
\end{equation}
The one-loop integral over $k_2$ can now be carried out with the procedure of
section~\ref{sc:def}.

The remaining two-loop integral $I^{(2)}_{\rm rem} \equiv I^{(2)}_{\rm reg} - \int \widetilde{dk}_1
\widetilde{dk}_2 \; G^{(2)}_{\rm glob}  - \int \widetilde{dk}_1
\widetilde{dk}_2 \; G^{(2)}_{\rm sub}$ is finite. After introducing Feynman
parameters for the $k_2$ integral, which are mapped onto a hypercube, and
shifting the $k_2$ loop momentum one arrives at an expression of the form
\begin{equation}
I^{(2)}_{\rm rem} = \int_0^1 dy_1 \dots dy_{n-1} \int \widetilde{dk_1}\widetilde{dk_2} \,
\biggl [ \frac{\tilde{C}_1(k_2)}{[\alpha k_1^2 + \beta k_2^2 - \tilde{A}]^n}
 + \frac{\tilde{C}_2(k_2)}{[\alpha k_1^2 + \beta k_2^2 - \tilde{A}]^{n-1}}
 + \cdots \biggr ],
\end{equation}
where $\alpha$, $\beta$ and $\tilde{A}$ are polynomials in the Feynman
parameters $y_1, ..., y_{n-1}$, and $\tilde{C}_i(k_2)$ are polynomials in $k_2$
and the Feynman parameters. Now one can perform the tensor reduction for $k_2$,
the $k_2$ loop integration, and---if necessary---the contour deformation in
analogy to eqs.~\eqref{eq:ten1}, \eqref{eq:int1}, and \eqref{eq:cd}.


\subsection{Numerical examples}

\begin{figure}[t]
\centering
\psfig{figure=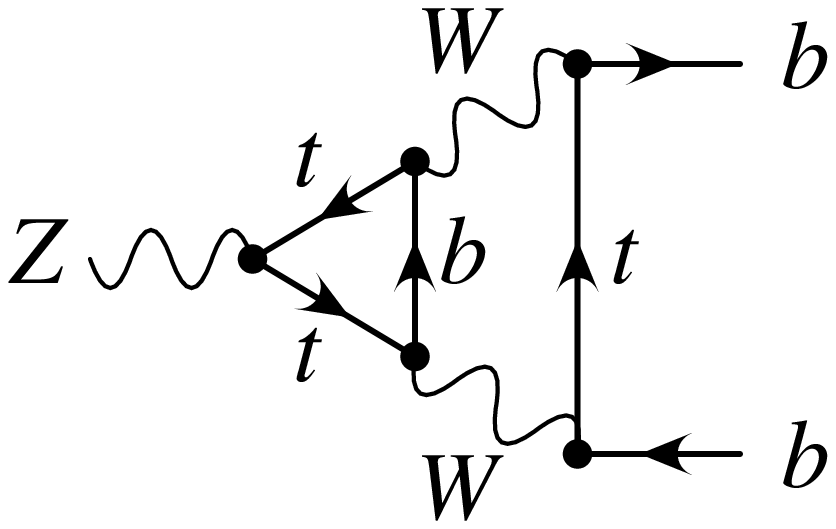, width=1.6in}
\vspace{-2ex}
\mycaption{Typical diagram for the two-loop correction to the $Zb\bar{b}$ vertex.
\label{fig:verE}}
\end{figure}

\paragraph{1.}
A typical example with global and subloop UV singularities is given by the diagram in
Fig.~\ref{fig:verE}, which contributes to the two-loop corrections to the
effective weak mixing angle of bottom quarks, $\sin^2\theta^{b\bar{b}}_{\rm
eff}$. Since this diagram does not have any physical cuts, no deformation of the
integration contour is needed and one can choose $\lambda=0$.
Tab.~\ref{tab:verE} shows the numerical results obtained with the algorithm of
the paper, using $10^6$ integration points of the {\sc
Cuhre} routine. The final answer is the sum of the global UV subtraction term, the subloop
UV subtraction term and the remaining finite two-loop integral. The integration
error is negligible except for the finite ${\cal O}(\varepsilon^{0})$ term.
The numerical integration takes about 13~min
to evaluate on one core of a {\sc Intel\textsuperscript{\textregistered} Xeon\textsuperscript{\textregistered}
X5570} processor with 2.93~GHz. For comparison,
the table also shows the result of Ref.~\cite{ew2}, which has been obtained with
the Bernstein-Tkachov (BT) method. The two results agree very well within
numerical integration errors.
\begin{table}
\centering
\renewcommand{\arraystretch}{1.2}
\begin{tabular}{lll}
\hline
 & \multicolumn{1}{c}{this work} & \multicolumn{1}{c}{Ref.~\cite{ew2}} \\
\hline
${\cal O}(\varepsilon^{-2})$ & $-$2.30183413 & $-$2.30183413 \\
${\cal O}(\varepsilon^{-1})$ & \phantom{$-$}5.07108758 & \phantom{$-$}5.07108758 \\
${\cal O}(\varepsilon^{0})$ & \phantom{$-$}8.326(1) & \phantom{$-$}8.3259 \\
\hline
\end{tabular}
\mycaption{Numerical results for the contribution of the diagram in
Fig.~\ref{fig:verE} to the effective weak mixing angle
$\sin^2\theta^{b\bar{b}}_{\rm eff}$, in units of $(\alpha/4\pi)^2$. The input
parameters are chosen as in Ref.~\cite{ew2}: $\MZ=1$, $\MW=80/91$, $\mt=180/91$.
The deformation parameter is set to $\lambda=0$. The numbers in the second
column have been obtained using $10^6$ integration points of the {\sc Cuhre}
integration routine, with the integration error given in brackets. The last
column shows the results from to Tab.~1 in Ref.~\cite{ew2} for comparison.
\label{tab:verE}}
\end{table}

\paragraph{2.}
Fig.~\ref{fig:ver2a} shows an example of a scalar diagram with both UV and IR
divergences. The soft subtraction in eq.~\eqref{eq:ss2} is used, but its
integrated form needs to be evaluated to order ${\cal O}(\varepsilon)$, due to
the presence of the UV singularity.
Numerical results obtained with this method are shown in the second column of
Tab.~\ref{tab:ver2a}, using $10^6$ integration points of the {\sc
Cuhre} routine, corresponding to a running time of 2.6~s. 
For this particular diagram an analytical result has been
obtained previously in terms of harmonic polylogarithms \cite{ver2},
which is also shown in the table for comparison.
\begin{table}
\centering
\renewcommand{\arraystretch}{1.2}
\begin{tabular}{lll}
\hline
 & \multicolumn{1}{c}{this work} & \multicolumn{1}{c}{Ref.~\cite{ver2}} \\
\hline
${\cal O}(\varepsilon^{-2})$ & $-0.43040894 + 1.40496295i$ 
			     & $-0.43040894 + 1.40496295i$ \\
${\cal O}(\varepsilon^{-1})$ & $-3.53105702$ & $-3.53105702$ \\
${\cal O}(\varepsilon^{0})$  & $-1.93471(1) - 2.08763(1)i$
			     & $-1.93471213 - 2.08762578i$ \\
\hline
\end{tabular}
\mycaption{Results for the numerical evaluation of the scalar diagram in
Fig.~\ref{fig:ver2a}. The second column shows the values obtained with the
method of this paper, using $m=1$, $s=5$, $\lambda=1$ and $10^6$ integration 
points of the {\sc Cuhre} algorithm. For comparison, the last column gives the
analytical result from eq.~(343) in Ref.~\cite{ver2}.
\label{tab:ver2a}}
\end{table}


\section{Summary}
\label{sc:sum}

This paper presents a procedure for numerically computing one- and two-loop
integrals using subtraction terms for the singular pieces of the integrand. A set
of subtraction terms for the removal of ultraviolet, soft and collinear
singularities has been described. The subtraction terms themselves can be easily
integrated analytically in dimensional regularization, while the difference
between the full loop amplitude and the subtracted contributions is finite and
can be evaluated numerically after taking the limit to four dimensions. The
approach is applied to general one-loop cases, as well as two-loop integrals
with global and subloop ultraviolet divergences but infrared divergences in
only one of the two subloops. The extension to overlapping infrared
singularities in both subloops will be discussed in a subsequent publication.

The finite numerical integral (after application of the subtraction terms and
expansion in the integration dimension) is evaluated in Feynman parameter space.
The integrand can still have singular points inside the integration regions if
the diagram has physical thresholds. These singularities are formally integrable but
are problematic for numerical integration routines, so that they must be avoided
by deforming the integration contour into the complex plane.

The usefulness of the proposed procedure has been demonstrated with several one-
and two-loop examples. One obtains an overall good convergence behavior of the
numerical integration, although problems can occur for integrals with pinch
singularities, which typically correspond to threshold configurations. It may be
possible to regularize the pinch singularities with additional finite
subtraction terms, but a detailed investigation of this matter is left for
future work.

The algorithm presented in this paper has been implemented in the public
computer program {\sc nicodemos} (available at {\tt http://www.pitt.edu/\~{}afreitas/}). It is based on
{\sc Mathematica} for symbolic manipulations and produces a {\sc Fortran}
executable for the numerical evaluation. The current
version 1.0 only contains one-loop functionality, but an extension to the
two-loop level is planned for future releases.


\section*{Acknowledgements}

This work has been supported in part by the National Science Foundation under
grant no.\ PHY-0854782.


\end{document}